\documentstyle[11pt,newpasp,twoside]{article}
\newcommand{\PSbox}[3]{\mbox{\rule{0in}{#3}\includegraphics{#1}\hspace{#2}}}

\begin{document}

\title{Evolution of Clustering and Bias in a $\Lambda$CDM Universe}
\author{Romeel Dav\'e}
\affil{Princeton University Observatory, Princeton, NJ 08544}
\author{Lars Hernquist}
\affil{Harvard-Smithsonian Center for Astrophysics, Cambridge, MA, 02138}
\author{Neal Katz}
\affil{Dept. of Astronomy, University of Massachusetts, Amherst, MA, 01003}
\author{David H. Weinberg}
\affil{Dept. of Astronomy, Ohio State University, Columbus, OH 43210}

\begin{abstract}
We determine the evolution from $z=3\rightarrow 0$ of the galaxy and
mass correlation functions and bias factor in a $50h^{-1}$Mpc
$\Lambda$CDM hydrodynamic simulation with $10h^{-1}$kpc resolution.
The mass correlation function grows with time, but the galaxy
correlation function shows little evolution and is well described by a
power law.  At early times, galaxies are biased
traces of mass, with bias being higher on smaller scales.  By $z=0$,
galaxies trace the mass, and the bias shows little scale dependence.
\end{abstract}


The correlation function of galaxies, $\xi_g$, is a standard observational
measure of structure in the Universe.  Understanding the
relationship between this observable and the correlation of the
underlying mass distribution, $\xi_m$, is crucial for relating
observations of galaxy clustering to predictions of cosmological
models.  The relationship between $\xi_g$ and $\xi_m$ is usually
characterized through the bias parameter $b\equiv
\sigma_g/\sigma_m$, where $\sigma$ is the rms fluctuation on some
specified scale.  Here we investigate the evolution of $\xi_g$, $\xi_m$ and
$b$ in a $\Lambda$CDM simulation.

We simulate a random $50 h^{-1}$Mpc cube in a $\Lambda$CDM universe,
with $\Omega_m=0.4$, $\Omega_\Lambda=0.6$, $H_0=65$, $n=0.95$, and
$\Omega_b=0.02h^{-2}$.  We use Parallel TreeSPH to advance $144^3$ gas
and $144^3$ dark matter particles from $z=49$ to $z=0$.  Our spatial
resolution is $10 h^{-1}$ kpc, and our mass resolution per particle is
$m_{SPH}=8.5\times 10^8 M_\odot$ and $m_{dark}=6.3\times 10^9
M_\odot$.  Using a 60-particle criterion for our simulated galaxy
completeness limit~[1] implies that we are resolving most galaxies with
$M_{baryonic}\ga 5\times 10^{10} M_\odot$.  We include star formation
and thermal feedback.  Galaxies are identified using Spline Kernel
Interpolative DENMAX.  At $z=3,2,1,$ and 0, we identify 929, 1781, 4138,
and 5264 galaxies, respectively.  Note that the majority of galaxies
form between $z=2$ and $z=1$ in this model.

We consider the 500 most massive galaxies at each redshift, representing
a constant comoving galaxy density of $n=0.004 h^3$Mpc$^{-3}$, comparable to
that of Lyman break galaxies and present-day $L_*$ galaxies.  The galaxy
correlation function $\xi_g$ at each redshift is shown as open circles
in Figure~1.  A power-law fit to these points is shown as the solid
line, and the best-fit values for the correlation length $r_0$ and
slope $\gamma$ are indicated in the upper right.  The mass correlation
function $\xi_m$ is shown as the dashed line.

\PSbox{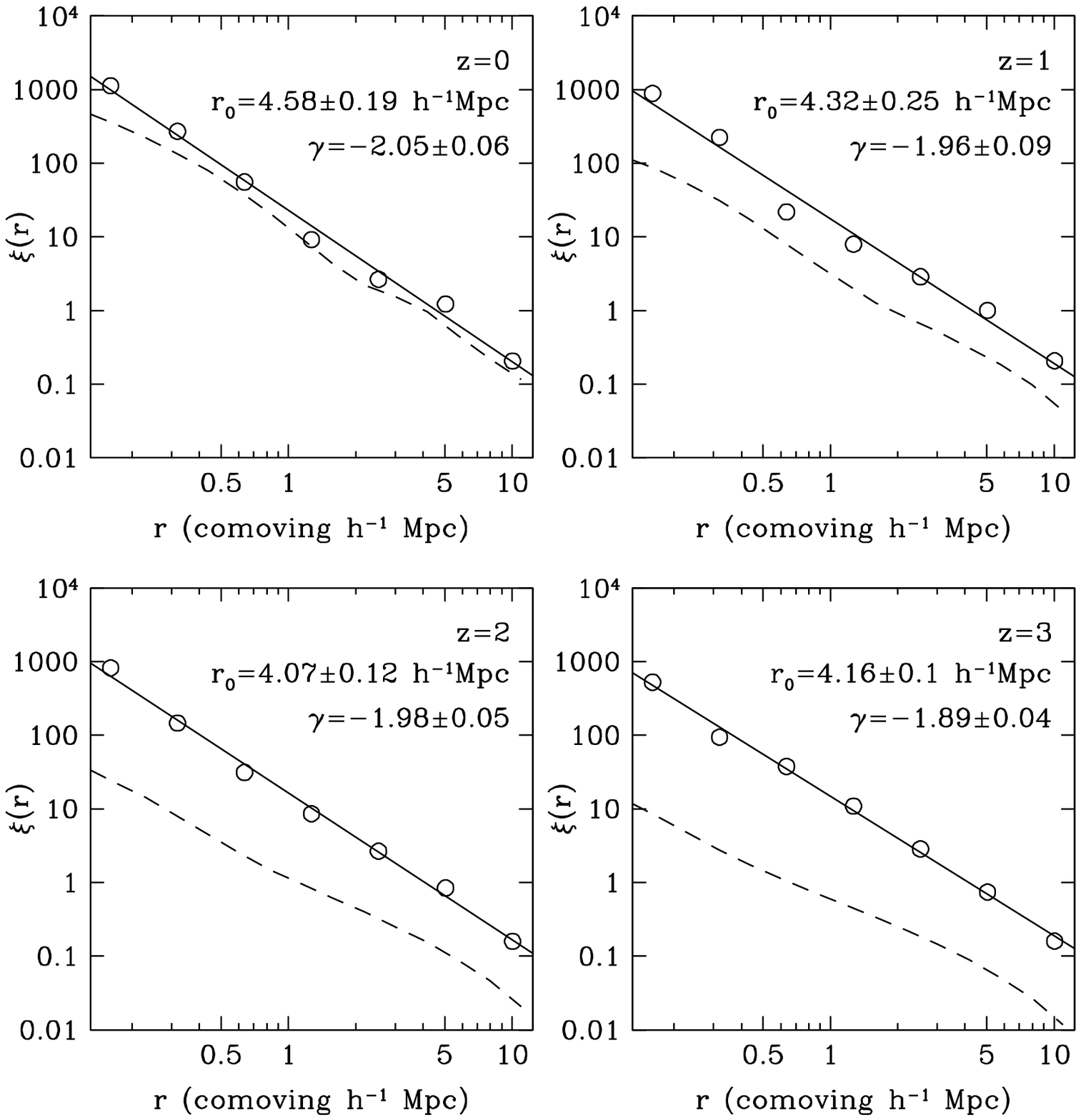 angle=0 voffset=-95 hoffset=-50 vscale=42 hscale=42}{3.0in}{3.0in}
{\\\small Figure 1 (left 4 panels):
Correlation function of galaxies and mass at various redshifts.
\label{fig: corr} }

\PSbox{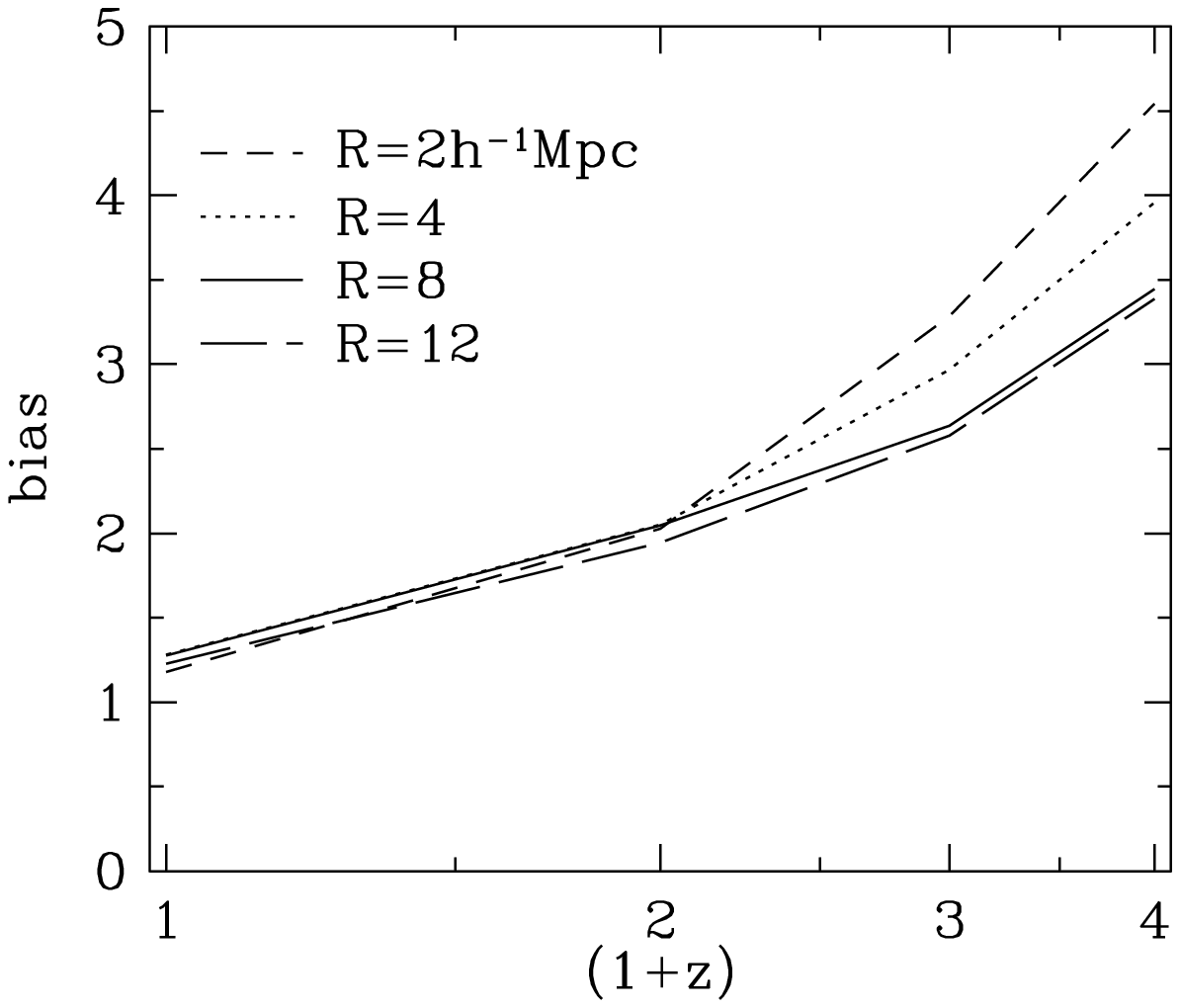 angle=0 voffset=-100 hoffset=150 vscale=42 hscale=42}{1.5in}{0in}
{\\\small Figure 2 (right panel): Galaxy bias on various scales as a function of redshift.
\label{fig: bias} }
\vskip0.1in
$\xi_g$ evolves little from $z=3$ to $z=0$, in agreement with previous
studies[2].  $r_0$ is roughly constant, with a slight increase to lower
redshifts.  $\gamma$ also evolves only slightly, with $\xi_g$ becoming
steeper at low redshift.  Conversely, $\xi_m$ increases substantially
with time.  The $z=0$ correlation function of this model is in good
agreement with observations.

The difference in evolution of $\xi_g$ and $\xi_m$ is reflected in the
evolution of $b(R)$.  We consider 4 scales, $R=2, 4, 8$ and
$12h^{-1}$Mpc.  The evolution of the bias parameter on these scales is
shown in Figure~2.  At early times, galaxies are highly biased tracers
of the underlying mass distribution, in agreement with previous
studies(see e.g. [3]).  This implies that Lyman break galaxies do not
trace the underlying mass distribution at high $z$.  For $R\la
8h^{-1}$Mpc, $b$ increases to smaller scales at high $z$.  The bias
factor declines with time on all scales, and shows little scale
dependence for $z\la 1$.  By $z=0$, $b\approx 1.2$ on all scales from
$R=2\rightarrow 12h^{-1}$Mpc, reflecting $\sigma_8=0.8$ chosen for our
simulation.

\end{document}